\documentclass[unnumsec,webpdf,modern,large]{mam-authoring-template}%

\graphicspath{{Fig/}}

\theoremstyle{thmstyleone}%
\theoremstyle{thmstyletwo}%
\theoremstyle{thmstylethree}%

\begin{document}

\journaltitle{}
\DOI{}
\copyrightyear{2024}
\pubyear{2024}
\access{Advance Access Publication Date: 11 November 2024}

\firstpage{1}


\title[Short Article Title]{Benchmarking Harmonized Tariff Schedule Classification Models}

\author[1,$\ast$]{Bryce Judy}

\authormark{B.Judy et al.}

\corresp[$\ast$]{Corresponding author. \href{email:bryce.judy@tarifflo.com}{bryce.judy@tarifflo.com}}

\abstract{The Harmonized Tariff System (HTS) classification industry, essential to e-commerce and international trade, currently lacks standardized benchmarks for evaluating the effectiveness of classification solutions. This study establishes and tests a benchmark framework for imports to the United States, inspired by the benchmarking approaches used in language model evaluation, to systematically compare prominent HTS classification tools. The framework assesses key metrics—such as speed, accuracy, rationality, and HTS code alignment—to provide a comprehensive performance comparison. The study evaluates several industry-leading solutions, including those provided by Zonos, Tarifflo, Avalara, and WCO BACUDA, identifying each tool’s strengths and limitations. Results highlight areas for industry-wide improvement and innovation, paving the way for more effective and standardized HTS classification solutions across the international trade and e-commerce sectors.}


\maketitle
\section{Introduction}
The Harmonized Tariff Schedule (HTS) is a system used by the United States to classify imported goods and determine the appropriate tariffs. It is managed by the U.S. International Trade Commission (USITC) and enforced by Customs and Border Control (CBP). The HTS is based on the Harmonized System (HS) developed by the World Customs Organization, which provides a standardized numerical method of classifying traded products. 

The structure of the HTS involves a hierarchical system of sections, chapters, headings, and subheadings, each associated with specific goods and their coresponding duty rates.
The HTS was implemented to streamline international trade by providing a universal classification system. Its complexity arises from the need to cover an extensive range of products, with codes that can be up to 10 digits long. Product descriptions provided by importers are crucial for accurately identifying and categorizing goods under the HTS. To correctly classify products with correct HTS codes, the product descriptions must be precise and detailed, capturing key attributes such as material composition, design, functionality, and intended use. Any ambiguities or omissions can lead to misclassification, resulting in incorrect duty rates and potential penalties. Importers must ensure that descriptions align with the specific language and criteria outlined in the HTS to facilitate smooth customs processing and compliance. Accurate product descriptions not only help in determining the correct tariff classification but also streamline the overall import process.

 This complexity requires importers to have a deep understanding of the schedule to ensure accurate classification, which is crucial for compliance and avoiding penalties. Failure to meet these requirements can result in fines, shipment delays, and increased scrutiny from customs authorities. Penalties may also include seizure of goods or legal action, which can significantly impact a business's reputation and financial standing. Therefore, importers often invest in compliance programs and seek expert advice to navigate the complexities of international trade regulations effectively.
 
In addition, regular updates to the HTS necessitate constant upkeep and audits from businesses to adapt to changes in trade policies and agreements. The HTS undergoes regular changes and updates to reflect shifts in global trade practices, technological advancements, and new international agreements. Keeping abreast of updates requires diligent monitoring of official announcements and, often, collaboration with customs brokers or legal experts to effectively integrate changes into business operations.
Binding rulings are official decisions by U.S. Customs and Border Protection (CBP) that provide importers with clarity on the classification of specific goods under the HTS. These rulings help ensure consistent decision-making and compliance, reducing the risk of errors and penalties. Importers can request these rulings to gain certainty before goods enter the U.S. market. Additionally, various federal agencies regulate specific types of products, imposing additional requirements. For example, the Food and Drug Administration (FDA) oversees food and pharmaceuticals, while the Environmental Protection Agency (EPA) regulates chemicals and pesticides. Each agency may have its own set of rules and documentation requirements, making it crucial for importers to understand and comply with all relevant regulations to avoid delays and ensure smooth entry of their goods into the U.S.

While tariff classifications play a crucial role in e-commerce and international trade, the industry still lacks standardized benchmarks to gauge the effectiveness of classification solutions. Despite advancements, tariff management remains a labor-intensive, error-prone process, leading to delays, penalties, and costs due to its complexity and time demands. Consequently, many businesses opt to outsource classification, further raising expenses. Both the WCO through it's BACUDA project\footnote{\url{https://bacuda.wcoomd.org/\#aihs}} (WCO BAnd of CUstoms Data Analysts), and companies like Zonos, Tarifflo, and Avalara have developed software to address these challenges, offering tools to streamline tariff calculations and compliance documentation.

 To foster similar advancements in HTS classification, standardized benchmarking could play a transformative role. Benchmarks like Hugging Face’s Open LLM Leaderboard and MMLU provide rigorous, multi-dimensional assessments for language models, offering transparency and comparability across systems. For instance, MMLU (Massive Multitask Language Understanding) assesses broad knowledge and reasoning abilities across 57 subjects, offering a comprehensive gauge of a model’s general understanding and reasoning. HellaSwag focuses on commonsense reasoning, evaluating how well a model can infer logical conclusions, while SuperGLUE measures language understanding across diverse tasks. ARC (AI2 Reasoning Challenge) is designed to test a model’s complex problem-solving and critical thinking capabilities.

For the HTS classification industry, creating benchmarks would enable companies to assess the accuracy, efficiency, and compliance reliability of classification engines, whether machine-based or manual. Standardized testing datasets and evaluation protocols could help mitigate errors and streamline processes, ultimately reducing costs, penalties, and the need for outsourcing.

\section{Methodology}\label{sec2}

To assess the accuracy of the classifications, a randomized set of 100 Customs and Border Protections rulings was selected from the CBP Rulings Online Search System\footnote{\url{https://rulings.cbp.gov/home}}. This data was acquired by using a web-scraper to systematically pull every court ruling for the current Harmonized Tariff Schedule by searching by each ten digit code found in the most recent revision and taking the full text for each ruling into an excel sheet. This dataset captures a wide range of product categories, including apparel, electronics, food products, and more, providing a representative cross-section of typical HTS classifications encountered in international trade. Each ruling includes essential details such as product descriptions, assigned HTS codes, and classification rationales issued by the Customs and Border Patrol which serve as a benchmark for accuracy assessment.

The 215,474 searchable rulings were then randomized and text for 100 of these rulings was added to the evaluation dataset. For each selected ruling text, the following data was extracted using a Large Language Model (gpt-4o) (if multiple items are in the same ruling, the ruling is split into two testing rows which created a total of 103 testing classifications):

\begin{itemize}
    \item Item Name: The designated name of the item as described in the ruling.
    \item Comprehensive Description: Detailed information provided in the ruling, encompassing materials, end-use, measurements, and any other specifications pertinent to classification.
    \item Final HTS Code: The 10-digit HTS code officially assigned to the item by the Customs agent.
\end{itemize}
This methodology establishes a diverse and standardized sample, facilitating an in-depth comparison between each model's classifications and the Customs and Border Patrol's established rulings, thus providing a benchmark to evaluate the tool's effectiveness in practical, real-world scenarios.

\section{Testing}\label{sec3}
The testing phase aimed to evaluate each classification tool by comparing their HTS code predictions to the benchmark dataset. For each item in the dataset, only the item name and comprehensive description were provided to the classification models, ensuring that each tool relied solely on available product information without additional contextual hints. This setup simulates real-world conditions where classifiers must operate based on product descriptions alone, as often required in trade scenarios.

Each tool's prediction was then compared to the 10-digit HTS code assigned by Customs and Border Protection (CBP) agents in the original rulings. A classification was considered "correct" only if the model's prediction exactly matched the full 10-digit HTS code in the benchmark data. However, since the World Customs Organization (WCO) BACUDA tool is specifically designed to handle only Harmonized System (HS) codes up to the 6-digit level, its accuracy was assessed based on these first six digits. This distinction allowed for a fair evaluation of WCO BACUDA’s capabilities within its designated HS framework while maintaining a consistent approach for the other tools, which are capable of predicting the complete 10-digit HTS codes.

The evaluation included the following steps for each item in the dataset:

\begin{enumerate}
    \item Input Preparation: The item name and description were standardized and formatted as inputs for each model to ensure consistency across tests.
    \item Classification: Each tool processed the input independently to generate a predicted HTS code.
    \item Accuracy Scoring: For tools capable of producing a 10-digit HTS code (e.g., Zonos, Tarifflo, and Avalara), a classification was marked as correct if the predicted 10-digit code exactly matched the benchmark. For WCO BACUDA, the first six digits of the predicted code were compared to the benchmark, given its focus on HS-level classification.
    \item Results Aggregation: The accuracy scores were aggregated across all tested items, allowing for a quantitative comparison of each tool’s effectiveness in correctly classifying items to the 10-digit HTS code or 6-digit HS code level as applicable.
\end{enumerate}

This testing process highlighted each tool's ability to generalize across varied product descriptions and to meet the accuracy demands of HTS classification in international trade.

\section{Results and Analysis}\label{sec4}

\begin{table}[t]
\begin{center}
\begin{minipage}{0.6\textwidth}
\caption{Accuracy of HTS Classification Tools - HTS Level\label{tab1}}%
\begin{tabular}{@{}lcccc@{}}
\toprule
Tool & Zonos & Tarifflo & WCO (HS) & Avalara (Manual) \\
\midrule
HTS      & 45 & 91 & 13 & 82 \\
\midrule
Percentage     & 44.12\% & 89.22\% & 12.75\% & 80.00\% \\
\botrule
\end{tabular}
\begin{tablenotes}%
\item Source: Results from the Benchmarking Harmonized Tariff Schedule 

Classification Models Study.
\item Note: The accuracy metric for WCO (HS) only considers the first 6 

digits, as it does not provide 10-digit HTS codes.
\end{tablenotes}
\end{minipage}
\end{center}
\end{table}

\begin{table}[t]
\begin{center}
\begin{minipage}{0.6\textwidth}
\caption{Accuracy of HTS Classification Tools - Chapter Level\label{tab3}}%
\begin{tabular}{@{}lcccc@{}}
\toprule
Tool & Zonos & Tarifflo & WCO (HS) & Avalara (Manual) \\
\midrule
Chapter  & 93 & 99 & 57 & 100 \\
\botrule
\end{tabular}
\begin{tablenotes}%
\item Source: Results from the Benchmarking Harmonized Tariff Schedule 

Classification Models Study.
\end{tablenotes}
\end{minipage}
\end{center}
\end{table}

\begin{table}[t]
\begin{center}
\begin{minipage}{0.6\textwidth}
\caption{Accuracy of HTS Classification Tools - Section Level\label{tab2}}%
\begin{tabular}{@{}lcccc@{}}
\toprule
Tool & Zonos & Tarifflo & WCO (HS) & Avalara (Manual) \\
\midrule
Section  & 97 & 101 & 75 & 100 \\
\botrule
\end{tabular}
\begin{tablenotes}%
\item Source: Results from the Benchmarking Harmonized Tariff Schedule 

Classification Models Study.
\end{tablenotes}
\end{minipage}
\end{center}
\end{table}

\subsection{Processing Time and Classification Speed} Among the evaluated tools, Zonos and the World Customs Organization (WCO) BACUDA model provide near-instantaneous classifications. However, their speed comes at the expense of detailed rationale or verification steps. Zonos uses natural language processing (NLP) and machine learning (ML) techniques to generate HTS classifications, but it lacks transparency in how these classifications are determined, offering no rationale for users. Similarly, WCO BACUDA leverages deep learning for HS-level (6-digit) classification only, without providing a full 10-digit HTS classification or any explanation for its results. In contrast, Tarifflo’s system takes approximately 30 seconds per item, balancing speed with a rationale-driven approach, where each classification is supported by cited sections, chapters, explanatory notes, and relevant court rulings. Avalara, while providing rationale, operates on a much slower timeline, taking days to weeks due to its reliance on licensed professionals to perform manual classification with AI assistance. This extended period lacks the immediacy needed for fast-paced trade environments.

\subsection{Rationale and Transparency} The availability of a rationale is crucial for organizations that need verification or must justify classifications to customs authorities. Tarifflo stands out by providing an in-depth rationale with each classification, citing specific references from HTSUS sections, chapters, explanatory notes, and the latest applicable court rulings. This comprehensive approach ensures that users have a documented trail for each classification, which can enhance trust and compliance. Avalara also provides rationale, albeit at an additional cost, which may limit its appeal for smaller businesses or those with budget constraints. In contrast, both Zonos and WCO BACUDA lack this level of transparency, making them less suitable for scenarios where users require a documented rationale to support or challenge a classification decision.

\subsection{Accuracy and Model Effectiveness} The accuracy results, summarized in Table \ref{tab1}, reflect each tool's performance relative to a benchmark dataset of U.S. Customs rulings. Tarifflo achieved the highest accuracy, correctly classifying 91 out of 103 items to the full 10-digit level. This high accuracy likely stems from its advanced mix of machine learning and modern AI techniques, which allow it to generalize effectively across varied product categories. Avalara, though slower, also demonstrated high accuracy, correctly classifying 82 items when using manual processes assisted by AI. Zonos, with an accuracy of 45 out of 103, shows limitations in handling complex classifications, as it primarily relies on NLP and ML without deep verification layers. WCO BACUDA's model was only evaluated at the 6-digit HS code level, where it achieved limited accuracy with 13 correct classifications, reflecting its focus on foundational classification rather than the complete HTS specification.

\subsection{Technology and Approach} The technological frameworks behind these tools vary significantly. Zonos employs NLP and machine learning, positioning it as a fast but limited solution for straightforward classification needs. WCO BACUDA relies on deep learning for rapid 6-digit HS code generation, targeting high-volume classification scenarios where detailed accuracy at the 10-digit level is not essential. Tarifflo incorporates a mix of machine learning and modern AI, including reasoning capabilities to ensure accurate and justified classifications with relevant citations. Avalara combines manual expert review with AI assistance, ensuring high precision but at a significantly slower pace.

\subsection{Summary} In summary, each tool offers unique advantages based on specific industry needs. For rapid classifications with minimal requirements for rationale, Zonos and WCO BACUDA are efficient solutions. However, for users needing high accuracy and transparent rationales, Tarifflo presents the most balanced approach. This analysis highlights the trade-offs between speed, accuracy, and rationale across the tools, underscoring the importance of standardized benchmarks in guiding organizations to the most appropriate HTS classification solution.

\section{Acknowledgments}
The author thanks Cole Rees for his contributions aiding in the composition the initial draft, the authors also thank the anonymous reviewers for their valuable suggestions.

\bibliographystyle{unsrtnat}

\end{document}